\documentclass[conference]{IEEEtran}
\IEEEoverridecommandlockouts
\usepackage{amsmath,amsfonts}
\usepackage{algorithm}
\usepackage[noend]{algpseudocode}
\usepackage{textcomp}
\usepackage{soul}
\usepackage{booktabs}
\usepackage[utf8]{inputenc}
\usepackage[T1]{fontenc}
\usepackage{microtype}
\usepackage{rotating}
\usepackage{graphicx}
\usepackage{paralist}
\usepackage{tabularx}
\usepackage{multicol}
\usepackage{multirow}
\usepackage{pbox}
\usepackage{enumitem}	
\usepackage{colortbl}
\usepackage{pifont}
\usepackage{xspace}
\usepackage{url}
\usepackage{tikz}
\usepackage{tabularx}
\usepackage{fontawesome}
\usepackage{lscape}
\usepackage{listings}
\usepackage{color}
\usepackage{anyfontsize}
\usepackage{comment}
\usepackage{soul}
\usepackage{multibib}
\usepackage{tcolorbox}
\usepackage{adjustbox}

\newcommand{\approach}{\emph{RETIWA}\xspace} 
\newcommand{\ie}{\emph{i.e.,}\xspace}
\newcommand{\eg}{\emph{e.g.,}\xspace}

\newcommand{\secref}[1]{Section~\ref{#1}\xspace}
\newcommand{\figref}[1]{Fig.~\ref{#1}\xspace}

\newcommand{\tabref}[1]{Table~\ref{#1}\xspace}

\newcommand{\smalltexttt}[1]{{\small \texttt{#1}}}

\definecolor{gray50}{gray}{.5}
\definecolor{gray40}{gray}{.6}
\definecolor{gray30}{gray}{.7}
\definecolor{gray20}{gray}{.8}
\definecolor{gray10}{gray}{.9}
\definecolor{gray05}{gray}{.95}

\newlength\Linewidth
\def\findlength{\setlength\Linewidth\linewidth
	\addtolength\Linewidth{-4\fboxrule}
	\addtolength\Linewidth{-3\fboxsep}
}
\newenvironment{rqbox}{\par\begingroup
	\setlength{\fboxsep}{5pt}\findlength
	\setbox0=\vbox\bgroup\noindent
	\hsize=0.95\linewidth
	\begin{minipage}{0.95\linewidth}\normalsize}
	{\end{minipage}\egroup
	\textcolor{gray20}{\fboxsep1.5pt\fbox{\fboxsep5pt\colorbox{gray05}{\normalcolor\box0}}}
	\endgroup\par\noindent
	\normalcolor\ignorespacesafterend}

\def\BibTeX{{\rm B\kern-.05em{\sc i\kern-.025em b}\kern-.08em
    T\kern-.1667em\lower.7ex\hbox{E}\kern-.125emX}}
\begin{document}

\title{Don't Reinvent the Wheel: Towards Automatic Replacement of Custom Implementations with APIs} 

\author{
\IEEEauthorblockN{Rosalia Tufano, Emad Aghajani, Gabriele Bavota}
\IEEEauthorblockA{\textit{SEART @ Software Institute,}
\textit{Universit\`a della Svizzera italiana, Lugano, Switzerland}\\
\{rosalia.tufano, emad.aghajani, gabriele.bavota\}@usi.ch}
}

\maketitle

\begin{abstract}
Reusing code is a common practice in software development: It helps developers speedup the implementation task while also reducing the chances of introducing bugs, given the assumption that the reused code has been tested, possibly in production. Despite these benefits, opportunities for reuse are not always in plain sight and, thus, developers may miss them. We present our preliminary steps in building \approach, a recommender able to automatically identify custom implementations in a given project that are good candidates to be replaced by open source APIs. \approach relies on a ``knowledge base'' consisting of real examples of custom implementation-to-API replacements. In this work, we present the mining strategy we tailored to automatically and reliably extract replacements of custom implementations with APIs from open source projects. This is the first step towards building the envisioned recommender.
\end{abstract}


\section{Introduction} \label{sec:intro}

Code reuse is a well-known practice aimed at improving both developers productivity and code quality \cite{Gaffney:1992}. 
There is evidence about the benefits of systematically reusing code, especially for what concerns a lower likelihood of having bugs in reused code \cite{Frakes:tse1996,Li:compsac2007,Haefliger:ms2008,Mohagheghi:tosem2008}. Such benefits arise when the reused code is not outdated \cite{Xia:imt2014} and follows good patterns of code reuse \cite{Kapser:emse2008}.
In such a context, Application Programming Interfaces (APIs) provide reusable functionalities that can be exploited by developers to (i) speedup the implementation of new features, and (ii) rely on well-tested implementations that have been possibly deployed in hundreds of client projects.
Despite these benefits, developers may not be aware of the availability of a specific feature implementation as an API offered by a third-party library, thus missing the opportunity of reusing it. To reduce such a risk, researchers proposed techniques aimed at recommending APIs given the coding context of the developer (\ie the code they are currently writing) \cite{Heinemann:suite2011,Thung:ase2016,Nguyen:tse2021}. While these tools could reduce the risk of reimplementing features offered in well-known libraries, such a scenario is still likely to happen, as also demonstrated by the results we present.

We describe our vision for \approach (\textbf{RE}placing cus\textbf{T}om \textbf{I}mplementations \textbf{W}ith \textbf{A}pis), a recommender to identify custom implementations (\ie code implemented from scratch) that can be replaced by third-party APIs. \approach is complementary to existing API recommenders \cite{Heinemann:suite2011,Thung:ase2016,Nguyen:tse2021}: The latter can recommend APIs while the developer is implementing. Differently, \approach comes into play once the custom code has been already implemented, automatically identifying it as a ``clone'' of a feature offered by well-known third-party APIs. 

This implies that \approach cannot save implementation effort, but can still avoid the maintenance of custom code that can be replaced with well-known and likely well-tested APIs, thus boosting code quality. 

We present the overall idea behind \approach and the first steps we made to build it. We started building a ``knowledge base'' featuring real replacements of custom implementations with APIs performed by developers in open source projects. We show the challenges behind such a task and the strategies we defined to address it. Such a knowledge base allows to identify  common ``replacement patterns'' that can be used to trigger custom implementation-to-API replacement recommendations.

\section{Envisioned Tool} \label{sec:tool}

\begin{figure*}
	\centering
	\includegraphics[width=0.95\linewidth]{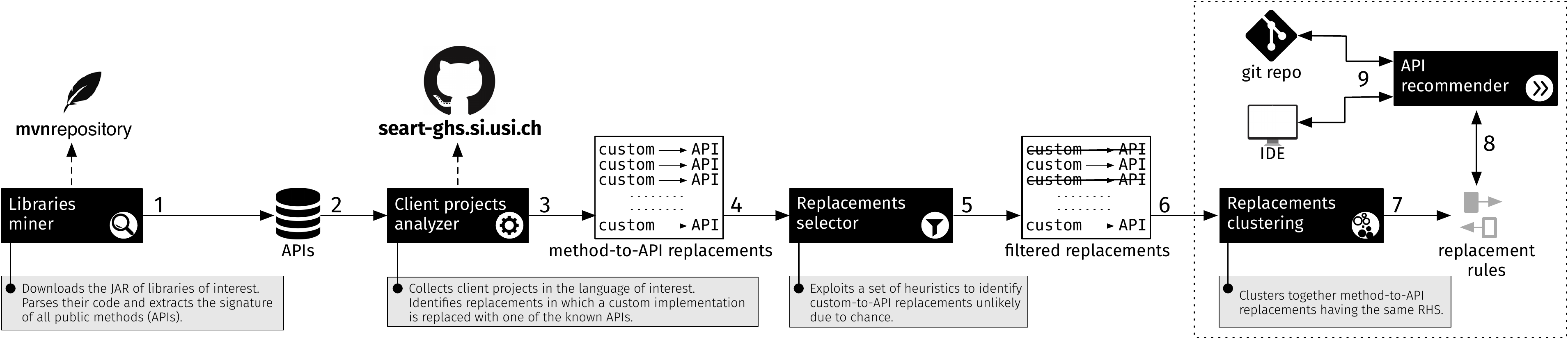}
	\caption{Overview of \approach Workflow}
	\vspace{-0.3cm}
	\label{fig:approach}
\end{figure*}

\figref{fig:approach} shows the main steps behind \approach. First, based on our preliminary tests, it is not realistic to reliably identify commits in which a custom implementation is replaced with an API without having a starting set of known APIs to support. Thus, our approach starts by mining software libraries with the goal of collecting a set of APIs (Step 1 in \figref{fig:approach}) through the \emph{Libraries miner} component. The libraries to support can be defined by the user. The set of mined APIs is then provided as input to the \emph{Client projects analyzer} (Step 2). The latter, given a collection of GitHub repositories, clones and analyzes the change history of the projects to identify candidate replacements where a custom implementation $c$ is replaced with one of the known APIs (Step 3). We indicate these custom implementation to API replacements as $c$ $\rightarrow$ $API$, with $c$ representing the lefthand side (LHS) and $API$ the righthand side (RHS). A single commit may contain multiple replacements, \ie several custom implementations are replaced with different APIs.

The parsing performed to identify such replacements may result in false positives. For this reason, the \emph{Replacements selector} applies a set of heuristics to exclude from the dataset instances likely to represent false positives (Step 5). The instances surviving such a filter are provided as input to the \emph{Replacements clustering} component (Step 6), which is in charge of grouping replacement instances characterized by the same RHS. These are code changes, possibly coming from different repositories, in which developers replaced a variety of custom implementations with the same $API$. The output of this step are the actual replacement rules $C$ $\rightarrow$ $API$, with $C$ being a set of custom implementations (Step 7). These rules can be used to recommend to developers the replacement of custom implementations with suitable APIs (Steps 8 and 9). 

In particular, a clone detector can be used to identify, either in a git repository or directly in the IDE, a component $p$ being similar to one of the clusters of custom implementations ($C$). 

Assuming the reliability of such a clone detector, the corresponding RHS (\ie the associated API) can then be recommended to the developer as a replacement for $p$.

We implemented the steps 1-6 shown in \figref{fig:approach} by instantiating \approach to the Java language and by working at method-level granularity: We look for replacement instances in the form $m$ $\rightarrow$ $API$, where $m$ is a Java method. In \secref{sec:study} we analyze the meaningfulness of the method-to-API replacements  that \approach was able to automatically identify, leaving the finalization of the recommender system (dashed part in \figref{fig:approach}) as future work. In the following, we provide details on the components we implemented and the design decisions we took.

\subsection{Libraries Miner}
\label{sub:miningLibs}
The first component is in charge of building a database of potential APIs that our approach may suggest as  replacement for an equivalent custom implementation. \emph{Libraries miner} currently supports the collection of APIs from Maven libraries.

Given a list of libraries of interest, \emph{Libraries miner} retrieves their source code by downloading and uncompressing the \smalltexttt{*-source.jar} for their latest release hosted on the \emph{Maven Central Repository} \cite{mvnrepo}. The set of public methods (APIs) in each library is then extracted using the \emph{Eclipse Java Parser} \cite{eclipse-jdt-core}. For each identified API we store the following information: package name, the path of the file from which it has been extracted, and its method signature. \emph{Libraries miner} also extracts all \smalltexttt{packages} defined in the source code of each library. Such information will be used to identify if \smalltexttt{import} statements in client projects refer to any of the target libraries.

\subsection{Client Projects Analyzer}
\label{sub:clientAnalyzer}
This component is responsible for cloning a set of given GitHub repositories and analyzing their change history with the goal of identifying potential replacements of a custom method with one of the APIs of interest. The analysis of each client project starts by creating a linear history of commits of the repository's default branch using the \smalltexttt{git log <default-branch> ---first-parent} command. Then, we iterate through all commits and perform the following steps on every two consecutive snapshots $s_{i}$ and $s_{i+1}$, where $s_{i}$ is the system's snapshot before the changes introduced by commit $c_i$ and $s_{i+1}$ is the snapshot after the changes introduced by $c_i$.

We start by extracting all \emph{method declarations} and \emph{method invocations} in the $s_{i}$ and $s_{i+1}$ snapshots. The method declarations represent all internal methods (\ie methods actually implemented in the system under analysis, excluding those imported from external libraries) existing in the system at a given snapshot. We indicate them with $D_{s_{i}}$ and  $D_{s_{i+1}}$. We use instead $I_{s_{i}}$ and  $I_{s_{i+1}}$ to indicate all method invocations existing in the two snapshots. These lists are extracted by checking out the corresponding snapshot (\ie\xspace\smalltexttt{git checkout <commit>}) and parsing the obtained Java files using SrcML \cite{srcml}. The main idea behind extracting these lists is to check in the following steps if the commit $c_i$: (i) replaced all method invocations to an internal method $m$ with invocations to a non-internal $API$; and (ii) deleted the implementation of the internal method $m$, replaced by the usage of $API$.

To perform this check, we start by running the command \smalltexttt{git diff $s_{i}$ $s_{i+1}$ $--$word-diff $--$unified=0 $--$\-ignore-\-all-\-space $--$F.java} for each file $F.java$ modified or renamed in $c_i$. This version of \smalltexttt{git diff} outputs pairs of \smalltexttt{[-oldcode-]} \smalltexttt{\{+newcode+\}} snippets, with the diff algorithm trying to match newly added code fragments (\ie\xspace\smalltexttt{newcode}) with deleted code ones (\ie\xspace\smalltexttt{oldcode}) when possible. We use such a command to identify invocations to $m$ replaced with invocations to $API$ in commit $c_i$. This is done by parsing the diff output using SrcML to match replaced method invocations. A few clarifications are needed on this step. First, we only focus on files modified or renamed in $c_i$ since we look for method invocation replacements which cannot happen in added or deleted files. Second, \smalltexttt{git diff} relies on heuristics to match deleted and added code fragments when possible, thus being a source of imprecisions in our approach. Third, similarly, SrcML is applied to parse fragments of code in the diff output, thus again resulting in imprecisions when matching method invocations. 

To exclude pairs $m$ $\rightarrow$ $API$ which are not of our interest (\eg $API$ is not an actual external API), we make sure that:

(1) $D_{s_{i}}$ contains $m$. This ensures that the method $m$ was an internal method declared in snapshot $s_i$. 
    
(2) The $m$ implementation in $s_i$ is non-empty (\ie it has a body) and does not invoke any method having the same exact name as $API$ (even with a different signature). This filtering step ensures that $m$ actually implemented something and, thus, could be a candidate to be replaced with an external API. 

\eject

Also, it excludes cases in which the developers were already using the $API$ (or a slightly different version of it taking different parameters), with $m$ acting as a wrapper for $API$. We are not interested in these cases since our goal is to recommend replacements of custom methods with APIs. However, if the developers are already aware of the existence of $API$, there is no reason for recommending it and probably they had a reason for not using it in the first place.
	
(3) $I_{s_{i+1}}$ contains $API$ (\ie the set of invocations in snapshot $s_{i+1}$ must contain the invocation to the candidate API), otherwise SrcML failed to recognize a true method invocation from the \smalltexttt{newcode} code fragment.
	
(4) There is no $m$ declaration in $D_{s_{i+1}}$ nor $m$ invocation in $I_{s_{i+1}}$. This indicates that (i) the implementation of the candidate custom method $m$ has been deleted from the system ($m$ $\not\in$ $D_{s_{i+1}}$); and (ii) no invocations to it exist anymore ($m$ $\not\in$ $I_{s_{i+1}}$).
	
(5) There is no method declarations matching $API$ in $D_{s_{i+1}}$, otherwise $API$ is a not an external API method.

The outcome of the aforementioned steps on commit $c_i$ is a set of candidate method-to-API replacements in form of  $m$ $\rightarrow$ $API$ for which (i) $m$ is a custom method that has been deleted in $c_i$; and (ii) $API$ is an external method that has been added in $c_i$ to replace all invocations to $m$. The \emph{Client projects analyzer} also makes sure that the $API$ is part of the libraries of interest provided as input to the \emph{Libraries miner}. This is done by verifying whether at least one of the added import statements in the files in which a $m$ $\rightarrow$ $API$ replacement happened matches with the list of packages extracted by the \emph{Libraries miner} for the libraries of interest. If this is not the case, the corresponding candidate replacement pair is removed. 

The final set of candidate replacement instances for each commit contains: (i) the GitHub repository owner/name; (ii) the commit sha; (iii) custom method information, including its signature and full implementation; (iv) information about the added API, including signature and potential libraries it belongs to based on the \smalltexttt{import} statements analysis; (v) file paths in which the replacements occurred; and (vi) the total number invocation replacements occurred in that commit (\ie the number of times an invocation to $m$ is replaced with $API$). 

\subsection{Replacements Selector}
\label{sub:selector}

This component is in charge of further filter out from the set of candidate replacement those instances likely to be false positives. Indeed, while implementing our approach, we noticed that two simple heuristics could be used to remove instances unlikely to be relevant for our goal. First, we exclude all replacement instances in which the custom method $m$ is either a getter, a setter, or a \smalltexttt{main} method. We do not see interesting scenarios in which it could make sense to recommend the replacement of these types of methods with APIs. 

Second, we conjecture that the number of $m$ $\rightarrow$ $API$ invocation replacements performed in a commit can be an indicator of how ``reliable'' is the method-to-API replacement we identified. 

\eject

Indeed, if we observe that in a given instance the internal method $m$ is removed and the several invocations to it that were present in the system are replaced with invocations to $API$, this (i) supports the idea that $m$ was a sort of ``utility'' method invoked in different parts of the code; and (ii) reduces the chances that the replacement is the result of a parsing error in the diff. We study how applying a threshold $t$ on the minimum number of call replacements impacts the reliability of the identified replacement instances (\secref{sec:study}).

%
\section{Preliminary Study} \label{sec:study}
The \emph{goal} of this study is to assess the feasibility of \approach in terms of the possibility to automatically identify changes in which custom implementations are replaced with APIs.
We aim at answering \textbf{RQ$_1$}: \emph{To what extent is it possible to automatically identify method-to-API replacements?} The automatic identification of $m$ $\rightarrow$ $API$ replacements is a pre-requisite for building \approach. RQ$_1$ looks at the precision of the approaches described in \secref{sec:tool}.


\subsection{Study Design}
The \emph{Libraries miner} collected 38 Apache \smalltexttt{commons} libraries by leveraging the MVN repository website \cite{mvnrepo}. These libraries can be identified as those having \smalltexttt{org.\-apache.\-commons} as maven group-id and have been downloaded and parsed as described in \secref{sub:miningLibs}. We used the SEART GitHub Search \cite{dabic2021sampling} to collect as client projects all non-fork Java GitHub repositories having at least 500 commits and 10 stars. These filters have been set in an attempt to exclude toy/personal projects. The \emph{Client projects analyzer} obtained 9,788 repositories as result of this search. However, only a fraction of these repositories (1,856) declared a dependency towards one of the considered libraries during their change history. These are the only projects from which we can expect useful data points for our study.

The analysis of these repositories, performed as described in \secref{sec:tool}, resulted in 337 candidate replacements which we manually analyzed to answer RQ$_1$. In particular, each $m$ $\rightarrow$ $API$ candidate replacement was inspected independently by two authors with the goal of classifying it as a true or false positive. To come up with such a classification, the inspector looked at (i) the diff of the commit on GitHub, (ii) a summary we created featuring all the invocations to $m$ that were replaced with an invocation to $API$, and (iii) the commit note. Conflicts, that arisen for 34 instances (10\%), have been solved by a third author not involved in the original classification. We report the precision of the identified instances (\ie the percentage of true positives among the candidate replacements) and discuss how it can be improved by increasing the minimum number $t$ of \emph{call replacements} (see \secref{sub:selector}). 


\subsection{Results Discussion}

\begin{table}[ht]
	\vspace{-0.3cm}
	\centering
	\caption{Manual analysis of 337 replacements identified by \approach\vspace{-0.3cm}}
        \label{tab:manual}
	\begin{tabular}{lrrr}
		\toprule
		\textbf{$t$} & \textbf{\# Instances} & \textbf{\# True Positives} & \textbf{Precision (\%)} \\ 
		\midrule
		$\geq 1$ & 337 & 165 & 48.9\%\\
		$\geq 2$ & 80 & 67  & 83.8\%\\
		$\geq 3$ & 46 & 39  & 84.8\%\\
		$\geq 4$ & 33 & 28  & 84.8\%\\
		$\geq 5$ & 25 & 23  & 92.0\%\\
		\bottomrule
	\end{tabular}
	\vspace{-0.2cm}
\end{table}

\tabref{tab:manual} reports the results achieved as output of our manual validation. The first column shows the threshold applied as additional filtering criterion to remove replacement commits not featuring at least $t$ call replacements. 

The first row (\ie $t \geq 1$) represents the scenario in which such a filter is not applied, since all instances \approach identified will have by construction at least one call replacement. 

As it can be seen, without any additional filtering the precision of the 337 identified replacements is limited to 48.9\%. By increasing the value of the $t$ threshold, the precision quickly increases, with a 83.8\% already achieved with $t \geq 2$ (\ie at least two invocations to the custom method $m$ have been replaced with the $API$ in the commit). Clearly, such an increase in precision has a cost in terms of true positive replacements that are excluded as having $t < 2$ (165-67=98 true positives are excluded). This is the usual recall \emph{vs} precision tradeoff that should be assessed based on how the envisioned recommender system will be built on top of this data. One option is also to provide developers with the possibility to decide the value of $t$: Higher values will result in less recommendations likely to be of high quality, while lower values, especially 1, will trigger more recommendations including, however, a higher percentage of false positives. In the following we discuss three concrete examples of replacements we found, while the whole dataset is available in our replication package \cite{replication}.

\begin{figure}[ht!]
	\centering
	\includegraphics[width=0.9\linewidth]{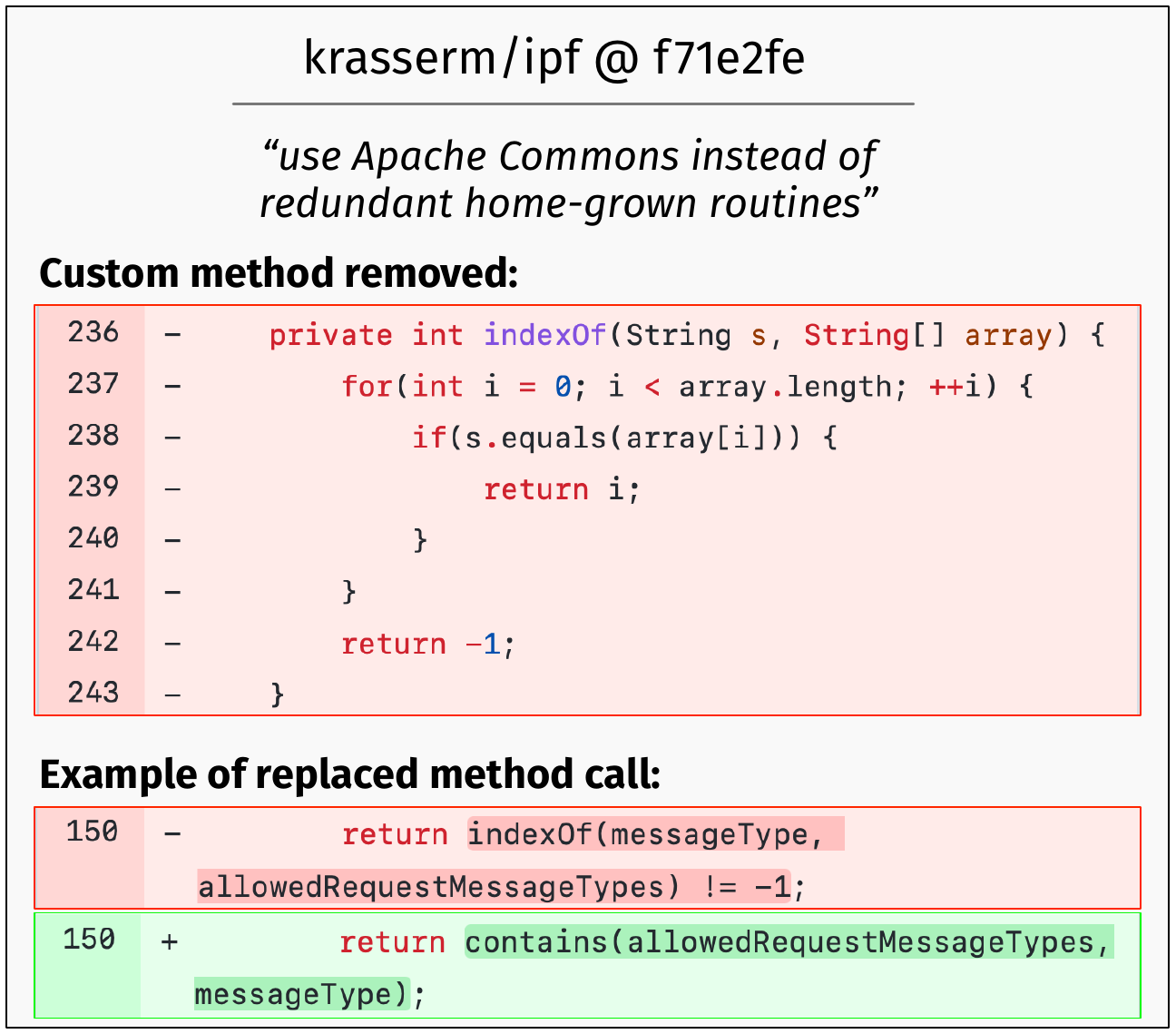}
	\vspace{-0.3cm}
	\caption{Replacing the custom \smalltexttt{indexOf} method with \smalltexttt{contains} API}
	\label{fig:ex1}
\end{figure}

In \figref{fig:ex1} the name of the GitHub repository and the commit we refer to are shown at the top. Following in \emph{italic} is the commit message used by the developer who explicitly indicates the aim of the commit of removing ``\emph{home-grown routines}'' in favor of APIs implemented in Apache commons. In the reported example the custom method \smalltexttt{indexOf} returns the index of a given \smalltexttt{String} in the \smalltexttt{array} provided as parameter or \smalltexttt{-1} if the \smalltexttt{array} does not contain the \smalltexttt{String}. 

This method is used in the system to check for the existence of the given \smalltexttt{String} in the \smalltexttt{array} as it can be seen from the replaced method call, in which it is used to return \smalltexttt{true} in case \smalltexttt{indexOf} returns a value \smalltexttt{!= -1}. The replacing API \smalltexttt{contains} already returns a \smalltexttt{boolean} thus also simplifying the locations in the code in which it is used instead of \smalltexttt{indexOf}.

\begin{figure}[ht!]
	\centering
	\includegraphics[width=0.9\linewidth]{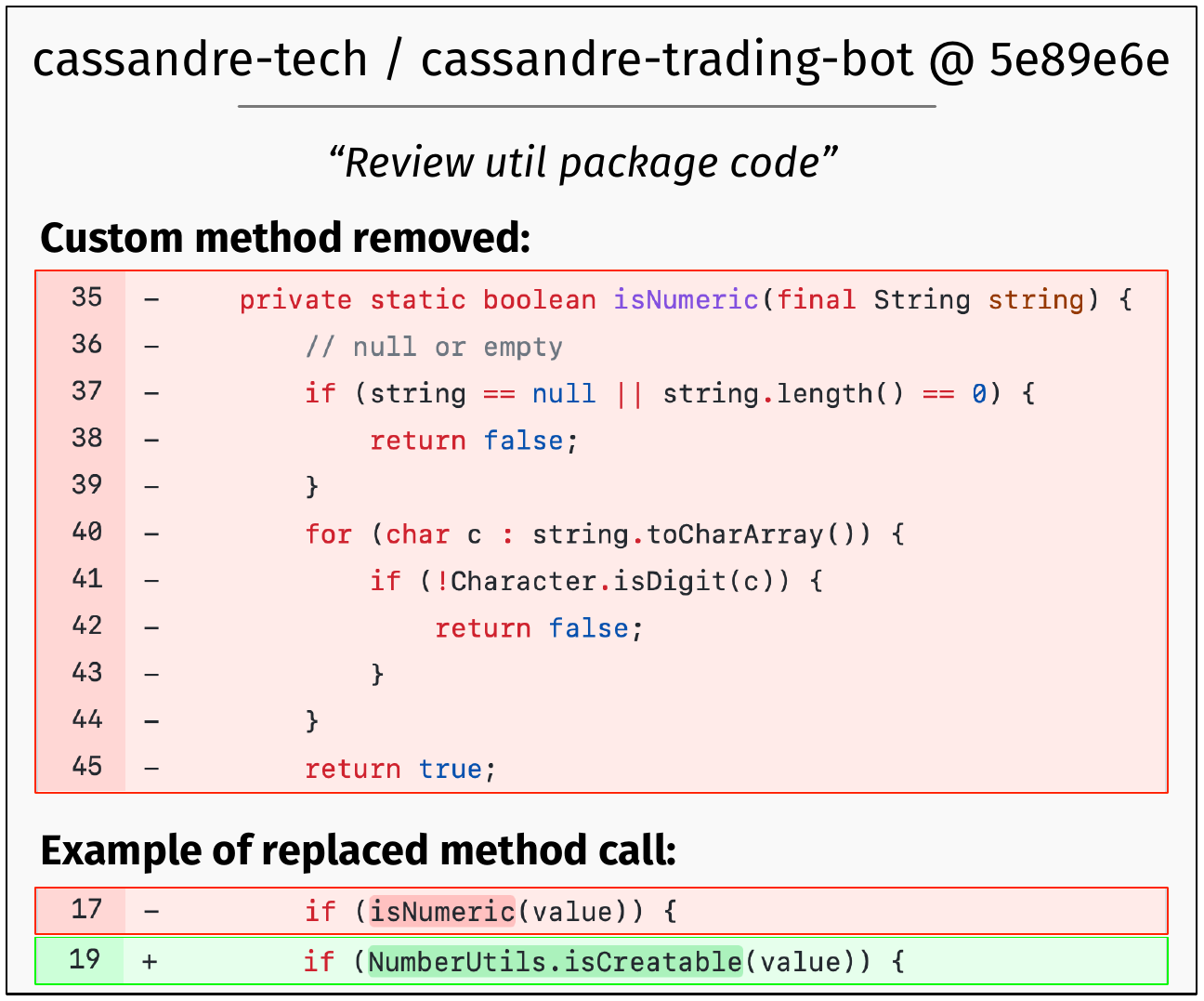}
	\vspace{-0.3cm}
	\caption{Replacing the custom \smalltexttt{isNumeric} method with \smalltexttt{isCreatable} API}
	\label{fig:ex2}
\end{figure}

\figref{fig:ex2} reports another example of replacement \approach identified. Differently from the previous commit, the commit message in this case does not allow to infer the presence of a $m$ $\rightarrow$ $API$ replacement without looking at the code diff. This is something we observed in the vast majority of commits in our dataset. Indeed, our initial idea was to match textual patterns in the commit messages to identify the candidate replacement commits as those containing \eg ``\emph{replaced custom [*] with [*]}''. However, such an approach is simply not an option due to the limited number of commits explicitly documenting these changes in their message. 

In this example, the replaced custom method is \smalltexttt{isNumeric}, which was in charge of verifying whether a \smalltexttt{String} provided as parameter was composed by only numbers. Invocations to such a method have been replaced by the \smalltexttt{NumerUtils.\-isCreatable} API which also takes as input a \smalltexttt{String} and, accordingly to its documentation, ``\emph{checks whether the String is a valid Java number}''.

\begin{figure}[ht!]
	\centering
	\includegraphics[width=0.9\linewidth]{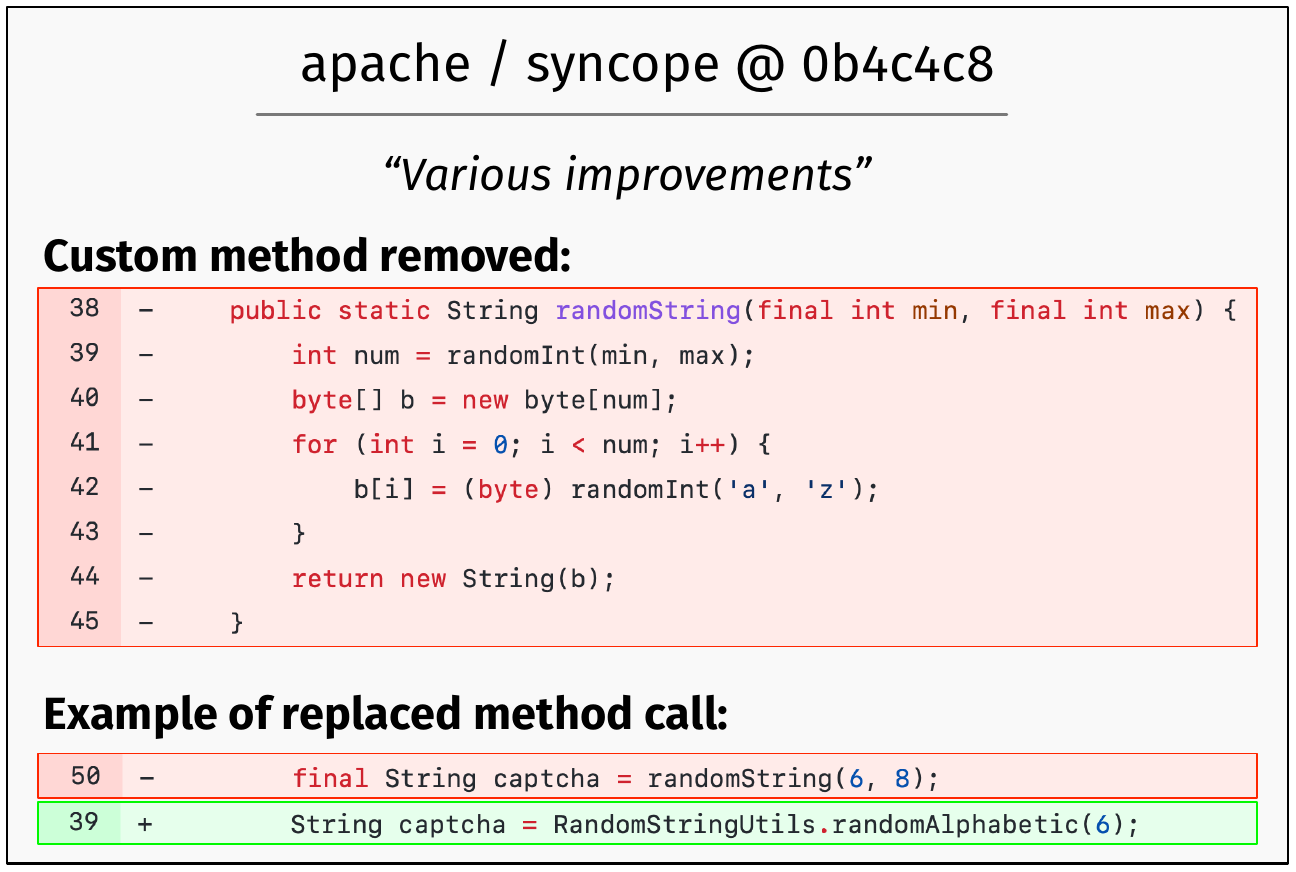}
	\vspace{-0.3cm}
	\caption{Replacing the custom \smalltexttt{randomString} method with \smalltexttt{randomAlphabetic} API}
	\label{fig:ex3}
\end{figure}

Finally, \figref{fig:ex3} depicts a commit, again characterized by a rather vague ``\emph{various improvements}'' message, in which the custom \smalltexttt{randomString} method used to generate random strings has been replaced with the \smalltexttt{randomAlphabetic} Apache API.

\begin{center}	
	\begin{rqbox}
		\textbf{Answer to RQ$_1$.} The automatic identification of changes replacing custom implementations with APIs is challenging but feasible. Indeed, the approach we presented was able to identify 337 of these commits with a precision of 48.9\% and, using specific filtering heuristics (\eg $t \geq 2$), the precision level can be substantially boosted to $>$ 80\%. More research is needed to optimize the recall \emph{vs} precision tradeoff. 
	\end{rqbox}	 
\end{center}

\subsection{Threats to Validity}
\textbf{Construct validity.} Our parsing procedure exploiting the output of the \smalltexttt{git diff} and SrcML may be subject to imprecisions when identifying the custom method invocations, the API invocations, and when mapping the API to the corresponding library through the analysis of the \smalltexttt{import} statements. Still, our goal was to investigate the feasibility of our approach and we are aware that better implementations based on full static code analysis of the involved snapshots can increase the parsing accuracy.

\textbf{Internal validity.} Subjectiveness in the manual analyses could have affected our results. To mitigate such a bias, when classifying the candidate replacement commits as \emph{true} or \emph{false positives}, two authors independently classified each comment, and a third author was involved in case of conflict. Despite such a process, imprecisions are still possible. The output of our manual analysis is publicly available for inspection \cite{replication}.

\textbf{External validity.} Our preliminary study focuses on a set of 38 Apache libraries and 1,856 Java client projects. Most of the steps behind \approach are independent from the target language, assuming the reimplementation of the low-level components such as the parser. Larger studies involving more diverse set of libraries are planned to corroborate our findings and designing the final version of \approach.

\section{Conclusions and Future Work} \label{sec:conclusion}
We presented our vision for \approach, an approach aimed at automatically identifying custom implementations that can be replaced by well-known third-party APIs. \approach takes as input a set of libraries of interest that are parsed to identify the APIs they contain. Then, a large set of client projects is mined to identify code changes in which developers replace a custom implementation with one of the known APIs. Heuristics are then applied to filter out false positives. By using the process we propose it is possible to automatically identify these replacements with a 48.9\% precision and such a precision can be boosted to $>$80\% by increasing the threshold $t$ (see \secref{sub:selector}). The replacement changes represent the basic data on top of which we plan to build the full recommender system depicted in \figref{fig:approach}. 

Towards this goal, our future work will span three main directions. First, we plan to explore alternatives for a more reliable code parsing. The obvious one is to perform a full parsing of the snapshots before and after each commit to identify custom code replaced with APIs. Such a process, while precise, is extremely expensive when applied on thousands of systems (\ie hundreds of thousands of commits to analyze).

Second, the combination of heuristics we adopt to identify replacement commits may be suboptimal and exclude several true positives. For example, \approach only selects as candidate replacement commits those in which a known \texttt{import} statement (\ie an \texttt{import} statement coming from one of the parsed libraries) is added in the commit. This is based on the assumption that the API usage implies the addition of the \texttt{import} statement. 

Third, once the identification of replacements is crystallized, we will focus on implementing the two remaining steps of \approach, namely the clustering and the triggering of recommendations, as described in \secref{sec:tool}.

\section*{Acknowledgment}
This project has received funding from the European Research Council (ERC) under the European Union's Horizon 2020 research and innovation programme (grant agreement No. 851720).  

\bibliographystyle{IEEEtranS}
\bibliography{main}

\end{document}